# Do Picardy thirds smile?

# Tonal hierarchy and tonal valence: explicit and implicit measures

Neta B. Maimon[a]*, Dominique Lamy[a,b] and Zohar Eitan[c]

[a] *The School of Psychological Sciences, Tel Aviv University, Tel Aviv, Israel*

[b]*The Sagol School of Neuroscience, Tel Aviv University, Tel Aviv, Israel*

[c]*Buchman-Mehta School of Music, Tel Aviv University, Tel Aviv, Israel*

*Corresponding author: POB 39040, Tel Aviv 69978, Israel. email:
netacello@gmail.com

# Do Picardy thirds smile?

# Tonal hierarchy and tonal valence: explicit and implicit measures

## Abstract

Western tonality provides a hierarchy of stability among melodic scale-degrees, from the maximally stable tonic to unstable chromatic notes. Tonal stability has been linked to emotion, yet systematic investigations of the associations between the hierarchy of melodic scale-degrees and perceived emotional valence are lacking. Here, we examined such associations in three experiments, in musicians and in non-musicians. We used an explicit task, in which participants matched probe tones following key-establishing sequences to facial expressions ranging from sad to happy, and an implicit speeded task, a variant of the Implicit Association Test. Stabler scale-degrees were associated with more positive valence in all experiments, for both musicians and nonmusicians. This notwithstanding, results significantly differed from those of a comparable goodness-of-fit task, suggesting that perceived tonal valence is not reducible to tonal fit. Comparisons of the explicit and implicit measures suggest that associations of tonality and emotional valence may rely on two distinct mechanisms, one mediated by conceptual musical knowledge and conscious decisional processes, the other – largely by non-conceptual and involuntary processes. The joint experimental paradigms introduced here may help mapping additional connotative meanings, both emotional and cross-modal, embedded in tonal structure, thus suggesting how "extra-musical" meanings are conveyed to listeners through musical syntax.

**Introduction**

Few topics in music cognition have been more widely studied than the cognition of Western tonality. A wealth of empirical research, using both implicit and explicit paradigms, has established tonality as a cognitive schema that orients the listener (for research surveys, see Krumhansl, 2004; Schmuckler, 2016; Shanahan, 2017). Much of that research validates the cognitive reality of a hierarchy of melodic scale-degrees, long suggested by music theorists (e.g., Rameau, 1722; Fétis, 1844; Schoenberg, 1978). In that hierarchy, the tonic note (scale degree 1) and, to a lesser extent, the other members of the tonic chord (scale degrees 3 and 5) are considered most stable and closural; the other diatonic scale degrees (2, 4, 6 and 7) are less stable and imply continuation to their more stable neighbors, 1, 3, or 5; the remaining five chromatic ("out of key") tones are the least stable, and strongly imply continuation to adjacent diatonic (within-key) notes[1]. This hierarchy was shown to significantly partake in shaping melodic expectancy (e.g., Schmuckler, 1989). Several studies also reported that following a tonal context, listeners' subjective evaluation of a tone's fit (e.g., Krumhansl & Kessler, 1982; Krumhansl, 1990), as well as the perceived level of musical tension associated with this tone (Eitan, Ben Haim, & Margulis, 2017) is correlated with the melodic scale-degree hierarchy. It is also consistent with pitch-classes' distribution in diverse tonal repertories, such that more stable scale-degrees tend to appear more frequently (e.g., Albrecht & Shanahan, 2013; Krumhansl & Schmuckler, 1990).

An increasing number of studies also examine the emotional connotations of tonality. They measure the impact of tonality and the expectations it generates on evoked emotion, that is, the emotions that the music evokes in listeners, as well as on perceived emotion, that is, the emotions that the music appears to express (for discussion of evoked vs. perceived emotion in music, see Kallinen & Ravaja, 2006; Kawakami et al., 2012). Most of these studies suggest that less stable tonal harmonies are associated with higher tension and arousal, and with more negative emotional valence. For instance, in Sloboda (1991), listeners exposed to tonal music reported experiencing shivers concurrently with the advent of unexpected and tonally unstable harmonies. Steinbeis, Koelsch, and Sloboda (2006) presented participants (musicians and non-musicians) with several versions of Bach chorales, while manipulating the degree of tonal stability of a single chord in each version. Participants reported elevated tension and overall emotionality for tonally unstable chords, compared to more stable ones; physiological measures, such as heightened skin conductance, corroborated these subjective reports. In a similar vein, Koelsch, Fritz, and Schlaug (2008) had nonmusicians listen to chord progressions in which the final chord could be either tonally stable (tonic resolution) or unstable (a chromatic substitute). Unstable progressions were rated as less pleasant than stable ones and were associated with increased amygdala activity. Finally, in Juslin et al. (2014), participants associated deviations from tonal expectancy, which generate unstable harmonic progressions, with negatively-valenced emotions, particularly anxiety, anger and irritation.

_______________________

1. For a more detailed account of tonal hierarchy, aimed at readers with no music theory background, see the introduction to Maimon, Lamy, & Eitan (2020).

Notably, the above studies involved harmonic, rather than melodic scale-degree hierarchy. Systematic investigations of the association of melodic scale-degrees with emotion are rare. To the best of our knowledge, no study has systematically investigated whether and how the entire gamut of melodic scale degrees in major and minor tonalities — the 12 chromatic tones — is associated with specific emotions (or, alternatively, with dimensions such as emotional valence or intensity). A comprehensive analysis of that relationship is essential to understanding how the scale-degree hierarchy established by both music theory and empirical music cognition studies may be related to the expression of emotion in tonal music. Such understanding may be a key to deciphering relationships between tonal syntax and emotion.

Two pioneering studies investigated emotional connotations of melodic scale degrees. In an informal survey, Huron (2006) asked 10 musicians to freely associate words and phrases related to different scale degrees in the major mode. Huron identified 7 semantic categories in participants' responses, one of which is emotional valence. Thus, for instance, participants associated stable diatonic degrees (1, 3, 5) with pleasure, love, or pleasantness, while unstable degrees – chromatic tones such as the raised tonic or raised subdominant – were perceived as "edgy" or "anxious." While these results are of interest, aspects of the study's methodology (e.g., free associations based on introspection, with no musical stimuli actually presented), the small sample and the fact that all participants were trained musicians, all limit its significance. In a follow-up to Huron's study, Arthur (2018) asked participants (musicians and non-musicians) to rate melodic scale degrees primed by different harmonic progressions on several bipolar adjective scales, including happy/sad, and examined the effects of harmonic priming and scale degree on these ratings. As in Huron's study, stable scale degrees (e.g., 1, 5) were associated with happier emotion, relative to unstable degrees.

Arthur's study, though addressing some of the issues in Huron's survey, is itself limited in substantial ways, insofar as the issue of correspondence between emotion and scale-degree hierarchy is concerned[2]. First, emotion (happy/sad) was not examined independently of other features: on each trial, participants responded to several different bipolar scales (e.g., tense/relaxed, move/stay, strong/weak) in the same way (by moving a slider). Responses in the happy/sad scale could thus be affected by responses to other scales. Second, like Huron (2006), Arthur's study considered only 10 out of the 12 tones, and only in the major mode. Third, it is difficult to interpret the effects of scale degree on happy/sad ratings in that study, because participants often opted out of rating this dimension (on 12.3% of the trials, as compared to 5% on average for other scales), and reported that valence rating, particularly for chromatic tones, was a particularly difficult task. Lastly, as responses were elicited directly, and relied on participants' introspection, the results may only reveal the explicit associations that exist between tonal hierarchy and emotion, particularly for musically-trained participants. Whether similar associations would arise if participants were not directly required to rate scale degrees remains unknown: finding that they do would suggest that the emotional associations of scale-degree hierarchy can also elicit unconscious, automatic responses.

---

2. We note that the association of emotion and scale degrees was not the principal issue of
   Arthur's study.

### *The present study*

This study investigated whether melodic tonal hierarchy – the hierarchy of stability and closure among melodic scale degrees, as postulated by music theorists and corroborated by cognitive and corpus studies – shows consistent, systematic correspondences with emotional valence. We examined such relationships for all 12 scale degrees within both major and minor contexts, which allowed us to determine whether there might be exceptions to such correspondences (e.g., instable scale degrees associated with positive valence, or vice versa). We employed both musically trained and untrained participants. Importantly, participants performed both an explicit (Experiment 1) and an implicit, speeded test (Experiment 2). Thus, both the roles of processes based on slow conscious reflection and those of faster, involuntary processes in associating tonal hierarchy and valence could be assessed and compared.

In the explicit tests (Experiments 1A, 1B), we applied a modification of Krumhansl's probe-tone paradigm (e.g., Krumhansl & Kessler, 1982). In that paradigm, participants hear a key-establishing "context element" on each trial (e.g., a cadence), which serves to prime a specific major or minor key, and then a single tone, the "probe." For each context element in a given key, the possible probes consist of the twelve pitch classes, with a different pitch-class presented in each trial. In Krumhansl's experiments, participants were asked to rate how well each probe fit with the context element that preceded it, on a 7-point scale. In the present study, this scale was replaced with 7 photographed faces of the same individual displaying a facial expression (taken from a validated database of emotional expressions), gradually ranging from sad to happy. Participants selected the face that best matched the probe tone, according to their subjective judgement.

The speeded implicit test (Experiment 2) was adapted from Parise and Spence's (2012) cross-modal version of the Implicit Association Test (IAT). In that paradigm, the stimulus set consists of two auditory and two visual stimuli. Two stimuli – one auditory and one visual – are assigned to the same response key in a given block of trials. Only one of the four stimuli is presented on each trial, and participants are required to respond as fast as possible using one of the two possible responses (different keys on the computer keyboard). The objective is to determine whether one visual stimulus is more strongly associated with one of the auditory stimuli (compatible stimulus) than with the other (incompatible stimulus). This can be inferred if participants' performance is better (faster RT, fewer errors) when compatible stimuli are assigned to the same response key. Here, we used a tonally stable and a tonally unstable stimuli for the auditory dimension, and two contrasting facial expressions (taken from the set used in Experiment 1) for the visual dimension. If the association of tonal stability and emotional valence can be elicited involuntarily, and is not necessarily dependent on conscious reflection, participants should respond faster and more accurately when the stable tone and the "happy" face were assigned to the one response key and the unstable tone and the "sad" face were assigned to the other response key, compared to when the alternative stimulus-response assignment was used.

Crucially, the IAT paradigm does not require any explicit judgement of how tightly the auditory and visual stimuli are associated. Such association – whether conscious or unconscious – is orthogonal to the task. Hence, any effects of congruence between tonal stability and emotional valence on the dependent variables (RT and accuracy) would reflect an unintentional, involuntary association between the two dimensions. Although the claim that the IAT paradigm taps automatic, wholly implicit, unconscious processes have been recently challenged, we believe that in the present context this paradigm should allow us to tap associations that are evoked rapidly, with little or no reflection – and thus substantially differ from the explicit associations examined, for instance, in Huron (2006) and Arthur (2018).

Based on earlier studies described above, we predicted that (1) tonal stability would be systematically associated with happier emotion: more stable scale degrees should be perceived as happier than less stable ones. (2) The association of tonal stability and valence should be reflected in both explicit (Exp1) and implicit (Exp2) tasks. (3) The association should be stronger for musically-trained participants – but mainly in the explicit task. We expected that musically trained subjects, in command of concepts associating musical features and emotion, should more strongly associate tonal stability and emotional valence than untrained participants. However, we predicted that training-based differences would emerge mainly on the explicit test, where the task allows conscious reflection, and not on the speeded IAT test, which taps rapid, involuntary responses. (4) Based on established emotional connotations of the major and minor modes (Parncutt, 2014; Smit et al., 2020), we expected that overall, major mode stimuli would be associated with more positive valence, compared to minor mode stimuli.

Finally, our main hypothesis (#1) notwithstanding, we evaluated whether our measure of "tonal valence" actually differs from established measures of perceived tonal stability, such as goodness of fit (GoF) ratings. For that purpose, we systematically compared the results from our explicit tests (Experiments 1a, 1b), with results of a previous GoF test using the same auditory stimuli (Maimon, Lamy, & Eitan, 2020), exploring differences between valence matchings and goodness-of-fit ratings both globally and with regard to scale-degrees of particular music-theoretical interest, such as the Picardy (raised) third.

## Exp. 1a: Matching scale degrees with facial expressions

### Materials and Methods

#### Sample size selection

Based on the study by Krumhansl and Kessler (1982), we calculated the sample size required to observe a significant effect of tonal hierarchy. Since they conducted multiple comparisons, we based our power analysis on the comparison between diatonic scale degrees and non-diatonic scale degrees, which had the smallest effect size. We conducted this analysis with G*Power (Erdfelder et al., 1996) using an alpha of 0.05, and a power of 0.80. We found the minimum required sample size to be four participants. As we examined possible interactions of this effect with musical training (musicians vs non-musicians) and mode (major vs. minor), the minimum sample size was 16.

*Participants*

Forty undergraduate and graduate students from Tel-Aviv University, 20 musicians (12 female, mean age = 28.65, SD = 6.52) and 20 non-musicians (15 female, mean age = 25.25, SD = 2.38) served as participants. The musicians had an average of 13.95 (SD = 4.27) years of musical experience (with a minimum of 7 years) and an average of 7.25 (SD = 3.51) years of music theory studies (with a minimum of 3 years). They all currently played and performed music. The non-musicians had an average of 2 (SD = 1.41) years of musical experience (with a maximum of 3 years in childhood), no music theory education, and none of them currently played music. All participants participated in the experiment for monetary credit (25$ for two 1-hour sessions, reported as Experiments 1a and 2).

*Apparatus*

The stimuli were presented on a computer with an Intel Core i7 CPU 920 processor, 2.67 GHz speed, and 2.98 GB RAM memory. Auditory stimuli were delivered through Sennheiser 210HD precision headphones. These were connected to the computer by a Terratec Producer, Phase 24 sound card. Stimulus loudness was measured through Brüel & Kjær type 2232 noise meter, measuring A-weighted dB. The computer screen was a 17-inch Lenovo LCD, with a screen refresh frequency of 85Hz and 1024 x 768-pixel resolution. The experiment was programmed and run using Matlab.

*Stimuli*

*Auditory stimuli.* Audio examples and graphical representations of the auditory stimuli are available in supplementary material A. In order to minimize pitch-height effects, all auditory stimuli were created using Shepard tones (Shepard, 1964). Shepard tones are tones with a specific pitch chroma (pitch note), sounded in 5 octaves simultaneously. A loudness envelope of 77.8-2349 Hz range is used, with gradual increase/decrease at its endings (Shepard, 1964). This loudness envelope generates tones with a clear pitch chroma but an ambiguous pitch height (i.e., the register in which a pitch is perceived – whether it is perceived in a higher or lower octave – is ambiguous, and determined contextually). All the stimuli were sounded in 71 dB.

Each trial consisted of a priming *element* followed by a *probe tone*. There were eight possible elements, including two element types, a triad and a cadence, each presented in four different keys. The triad – either a major or a minor chord – was presumably (based, e.g., on results in Krumhansl & Kessler, 1982) perceived as a tonic chord (the first, third and fifth scale degrees played simultaneously). The cadence was a perfect authentic cadence, featuring the chord sequence IV-V-I, which signifies strong closure in tonal theory. Each element was presented in two major and two minor keys (G major, G minor, D-Flat major and D-Flat minor). There were 12 possible probes: the 12 tones of the chromatic scale (12 notes from C to B). On each trial, an element was played for .5 sec when it was a triad, and for .5 sec for each chord followed by a silence of .25 sec when it was a cadence. Then, following a silence of 1 sec, a probe tone was played for .5 sec.

The experiment consisted of 18 blocks. The first two blocks served as practice. The elements for these blocks were randomly drawn from the eight possible elements and consisted of a cadence and a triad, in either G or D Flat (counterbalanced across subjects). They were followed by 16 experimental blocks, two for each of the eight possible elements, randomly mixed. The same element was used throughout any given block of trials. Each block consisted of 14 consecutive trials. The first two trials served as practice, using two randomly selected probes. In the following 12 experimental trials, the 12 chromatic tones, randomly mixed, were used as probe tones.

*Visual stimuli.* The visual stimuli consisted of seven faces with emotional valence ranging from sad to happy. We used photographs of veridical faces taken from the Data Organization of BU-3DFE Database. This database contains 100 individuals performing six prototypic expression (*happiness, disgust, fear, angry, surprise* and *sadness*), at four levels of intensity, plus one neutral expression. In the present experiment, we selected four Caucasian individuals (two women), and generated 4 7-level continua by selecting the first three levels of sadness and the first three levels of happiness and adding the neutral expression as the middle level (see example in figure 1).

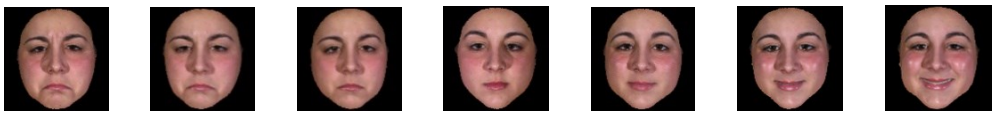

Figure 1. Example of one (out of four) gradient scales of emotional faces created with the BU-3DFE Database.

*Procedure*

Participants underwent two sessions, the unspeeded-task session and the speeded IAT task session, which were administered 7 to 14 days apart (M=8.62). Task order was counterbalanced across participants. The unspeeded session is reported here (Experiment 1a) and the speeded IAT session is reported as Experiment 2 (pp.27-30).

For each block, one continuum out of the four (consistent throughout the experiment, with the saddest face either presented on the left and the happiest face on the right, or vice-versa, counterbalanced between participants) appeared on the screen throughout the entire block. After listening to the element followed by the probe, participants were asked to mark the face, which, according to their subjective judgment, matched each probe tone best. They were required to click the corresponding face with the cursor.

The experimenter underscored the subjective and intuitive character of the task. She instructed participants to choose whatever strategy they felt was the most suitable to perform the task, yet to be consistent and use the same strategy throughout the experiment.

After each block, participants were asked about their confidence in the judgments they had provided. The text "How confident were you with the ratings of the last block?" appeared on the screen and participants provided a 1-7 rating by pressing the appropriate numeral on the upper row of the keyboard. Then, a 10-sec. white noise (71 db) was heard, followed by a count down from 5 to 1. Participants then proceeded to the next block by pressing any key, after a self-pace break. All written and oral instructions were presented in Hebrew.

**Results**

*Main analysis*

Preliminary analyses showed no significant effect involving element types (i.e., cadence vs. triad). The data were therefore collapsed across these variables. All pairwise comparisons were conducted using Bonferroni corrections.

A repeated-measure analysis of variance (ANOVA) was conducted, with musical training (musician vs. non-musician) as a between-participants variable and tonal stability and mode (major/minor) as within-participant variables. The tonal stability variable included three categories: *Stable diatonic* (the tonic chord members – the 1$^{st}$, 3$^{rd}$, and 5$^{th}$ degrees of the scale; that is, tones 1, 5, 8 of the 12 chromatic tones in major and tones 1, 4, 8 in minor), *Unstable diatonic* (2$^{nd}$, 4$^{th}$, 6$^{th}$ and 7$^{th}$ degrees of the scale: tones 3, 6, 10, 12 in major and tones 3, 6, 7, 11, 12 in minor) and *chromatic tones* (non-scale degrees: tones 2, 4, 7, 9, 11 in major and tones 2 ,5 ,7, 10 in minor). A summary of the results is presented in Figure 2.

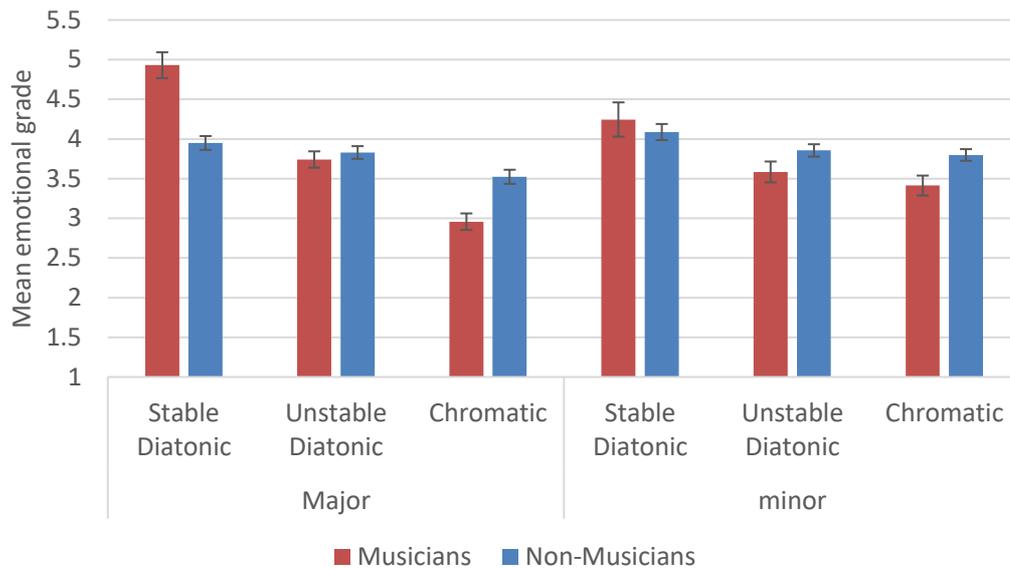

Figure 2. Mean emotional faces matchings in Experiment 1a as a function of tonal stability categories, in major (right panel) and minor (left panel) modes and for musicians (blue bars) and non-musicians (red bars). 1=happiest face, 7=saddest face. Bars denote standard errors.

The main effect of tonal stability was significant, $F_{(2,74)}=43.30$, p<.001, $\eta_p^2=.53$. Pairwise comparisons revealed that emotional faces matchings were happier for the stable diatonic than for the unstable diatonic category, $t_{(1,38)}=4.95$ p<.001, and these were happier than emotional faces matchings for the chromatic-tone category, $t_{(1,38)}=5.1$ p<.001. There were no other significant main effects. All the two-way interactions were significant: tonal stability interacted with mode, $F_{(2,74)}=15.8$, p<.001 $\eta_p^2$=.3, and with musical training, $F(2,74)$=16.19, $p$<.001, $\eta_p^2$=.3, and mode interacted with musical training, $F(2,74)$=6.58, $p$=.015 $\eta_p^2$=.15. These interactions were modulated by a significant three-way interaction between tonal stability, musical training and mode, $F(2,74)$=9.38, $p$=.001, $\eta_p^2$=.2. To clarify this interaction, we conducted separate ANOVAs for musicians and for non-musicians.

*Musicians*

The main effect of tonal stability was significant, $F(2,38)$=34.59, $p$<.001, $\eta_p^2$=.65. Pairwise comparisons revealed that emotional faces matchings were happier for the stable-diatonic than for the unstable-diatonic category, $t(1,19)$=5.21, corrected $p$<.001, and these were happier than emotional faces matchings for the chromatic-tone category $t(1,19)$=4.62, corrected $p$<.001. The main effect of mode was not significant, $F(2,38)$=2.53, $p$=0.13, $\eta_p^2$=.117, but interacted with tonal stability, $F(2,38)$=15.54, $p$<.001 $\eta_p^2$=.45, indicating that the differences in musicians' emotional faces matchings between tonal stability categories were wider within the major than within the minor mode. Simple effects analyses revealed that within the major mode, the difference between stable diatonic and unstable-diatonic categories was significant, $t(1,19)$=7.71, corrected $p$<.001, and so was the difference between unstable-diatonic and chromatic tones categories $t(1,19)$=6.44, corrected $p$<.001. Within the minor mode, the difference between stable- and unstable-diatonic categories was only marginally significant, $t(1,19)$=2.62, corrected $p$=0.051, and the difference between the unstable diatonic and the chromatic tones categories was not significant, $t(1,19)$=1.76, corrected $p$=0.285.

*Non-musicians*

The main effect of tonal stability was significant, $F(2,36)$=11.01, $p$<0.001, $\eta_p^2$=.38. Pairwise comparisons revealed that emotional faces matchings tended to be happier for the stable-diatonic than for the unstable-diatonic category t(1,18)=2.6 corrected p=0.055, and these were significantly happier than emotional faces matchings for the chromatic-tone category, t(1,18)=2.9, corrected p=0.028. The main effect of mode was significant, with (surprisingly) happier emotional faces matchings for the minor mode than for the major mode F(2,36)=4.63, p=.045, $\eta_p^2$=.21. The interaction between tonal stability and mode approached significance, F(2,36)=3.17, p=.054, $\eta_p^2$=.15, indicating that within the major mode, the difference between the unstable-diatonic and the chromatic tones was significant, t(1,18)=4.1, p=.001 and the difference between stable-diatonic and unstable-diatonic categories was not, t(1,18)=1.16, p=.26, whereas within the minor mode, the difference between the stable-diatonic and unstable diatonic was significant, t(1,18)= 3.34, corrected p=.004, whereas the difference between unstable-diatonic and chromatic tones categories was not significant, t<1.

*Confidence ratings*

The overall mean confidence rating was M=4.74 (SD=1.61). An ANOVA with musical training (musicians/non-musicians) as a between-participants variable and mode (major/minor) as a within–participant variable was conducted on the mean confidence ratings obtained. No effect reached significance: musical training, M=5.29, SD=1.14 vs. M=4.68, SD=0.87 for musicians and non-musicians, respectively, F(1,37)=3.48, p=.07; mode, M=4.78, SD=1.62 vs. M=4.69, SD=1.58 for the major and minor modes, respectively; musical training X mode, F<1.

**Discussion**

In general, results of Experiment 1a revealed a correspondence between tonal stability and emotional valence: happier faces were matched to more stable tonality probes. Though this association was significant for both musicians and non-musicians, it was strengthened by musical training, which suggests that it relies, at least in part, on conceptual musical knowledge.

However, the scalar, linear ordering of the visual stimuli (faces ordered from either happy to sad or from sad to happy) might have encouraged participants to construe the task as a ranking task, without specifically relying on their perception of the emotional expressions conveyed by the faces. If so, the schema relating probe-tones and valence along a shared scale would reflect, at least in part, an artifact of the experiment's design. In Experiment 1b we addressed this issue by replicating the emotional faces matching task of Experiment 1a, while ordering the faces randomly, rather than as a continuum between the happiest and the saddest facial expressions.

## Experiment 1b

### *Materials and Methods*

#### *Participants*

Twenty undergraduate students from Tel-Aviv University served as participants, 10 musicians (2 female, mean age = 22.9, SD = 2.84) and 10 non-musicians (5 female, mean age = 24.5, SD= 4.03). The musicians had an average of 14.6 (SD= 3.94) years of musical experience (with a minimum of 9 years) and an average of 8.8 (SD= 5.18) years of musical theory studies (with a minimum of 5 years). They all currently played and performed music. The non-musicians had an average of 1.375 (SD 0.75) years of musical experience (with a maximum of 3 years in childhood), no music theory education and none of them currently played music. None of the participants had taken part in Experiment 1a. All participants were paid 10$ per hour for their participation.

#### *Apparatus, stimuli and procedure*

The apparatus, stimuli and procedure were similar to those of Experiment 1a, except for the following differences. The seven faces were presented in random order rather than as a monotonic progression between one facial expression of emotion to the other. A specific random order and a specific facial identity remained constant in each block and both changed between the blocks. Participants were notified that ordering and identity changes would occur during the experiment. Unlike in Experiment 1a, participants were not required to rate their confidence.

### *Results*

### *Main analysis*

We conducted a repeated-measures ANOVA with musical training (musician vs. non-musician) as a between-participants variable and tonal stability (stable diatonic, unstable diatonic, and chromatic tones) and mode (major/minor) as within-participant variables. All pairwise comparisons were conducted using Bonferroni corrections.

The mean valence (emotional faces) matchings are presented in Figure 3. The main effect of tonal stability was significant, F (2,36) = 28.02, p<.001, $\eta_p^2$=.61. Pairwise comparisons revealed that participants matched happier faces with stable diatonic tones, compared to both unstable diatonic tones (t(1,19)=5.45, p<.001) and chromatic tones (t(1,19)=5.29, p<.001); there was no significant difference between unstable diatonic and chromatic tones, t(1,19)=1.98, p=0.161. Tonal stability interacted with musical experience, F(2,36)=4.94, p=.013, $\eta_p^2$=.215, and with mode F(2,36)=9.27, p=.001, $\eta_p^2$=.34. These interactions were modulated by a higher-order interaction between tonal stability, mode and musical experience, F(2,36)=5.11, p=.011, $\eta_p^2$=.221.

Separate ANOVAs for musicians and non-musicians revealed that while mode (major/minor) affected the impact of tonal stability on emotional face matching by musicians, F(2,18)=13.94, p<.001 $\eta_p^2$=.61, this was not the case for non-musicians, F<1. Specifically, for musicians, matchings were happier for stable than for unstable diatonic tones for both major and minor modes, t(1,9)=4.53, p=.004 and t(1,9)=3.46, p=.022, respectively; Musicians' matchings were also happier for the unstable diatonic than for the chromatic tones, but only in the major mode, t(1,9)=3.87, p=.004. For non-musicians, matchings were also happier for stable than for unstable diatonic tones in both major and minor modes, t(1,9)=3.06, p=.041 and t(1,9)=2.87, p=.056, respectively. However, for non-musicians (unlike musicians) no difference was found between unstable diatonic and chromatic tones, in either major or minor (p >.05).

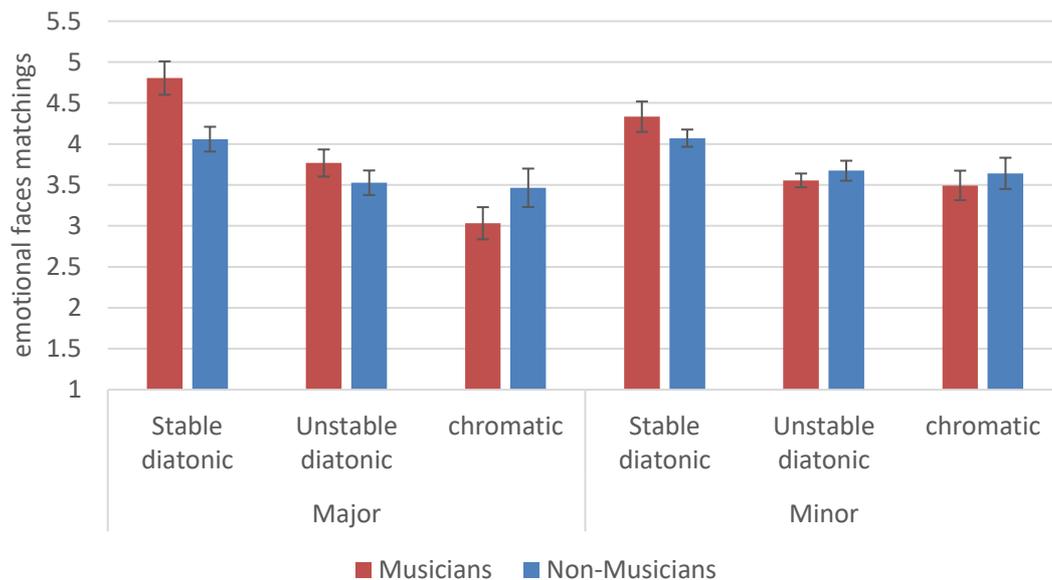

Figure 3. Mean valence (emotional faces) matchings in Experiment 1b (with emotional faces presented in randomized order), as a function of tonal stability categories, in major (right panel) and minor (left panel) modes and for musicians (red bars) and non-musicians (blue bars). 1=happiest face, 7=saddest face.

***Discussion***

In Experiment 1b, by randomizing the order of the emotional faces (rather than presenting them as a monotonic progression, as in Experiment 1a), we prevented participants from relying on explicit top-down processing. Results of this experiment replicated the main finding of Experiment 1a, demonstrating correspondence between tonal stability and emotional valence. We also replicated the modulation of this correspondence by musical training, as the effect of tonal stability was larger within musicians.

### *Comparing Tonal Valence and Goodness-of-fit measures*

Results of Experiments 1a and 1b indicate that tonal stability is significantly associated with tonal valence: stabler scale degrees were associated with happier facial expressions. Yet, as a devil's advocate may suggest, is that so because we merely replicated, with insubstantial variation of experimental procedure, established measures of perceived tonal stability, rather than examine an emotional dimension? Specifically, one may suggest that the present experiments are simply variants of goodness-of-fit (GoF) probe-tone measures (e.g., Krumhansl & Kessler, 1982), substituting the original rating task with an equivalent matching task.

Here we present an analysis suggesting that perceived tonal valence, as examined in the above experiments, and perceived tonal stability, as revealed by GoF measures, are different, distinct features. To examine the relationships between these features we compared mean valence ratings in Experiments 1a and 1b (combined) with mean GoF ratings in a recent probe-tone experiment (Maimon et al., 2020), which used identical auditory stimuli with a similar participants group (n=40, 20 musicians). We performed a repeated-measures ANOVA on the data of these experiments, with tonal stability (Stable diatonic, Instable diatonic, Chromatic) as a within-subject variable, and rating type (GoF, Valence) and musical experience (Musician, Non-musician) as between-subjects variables. Mean rating was the dependent variable.

Results (Figure 4 and Table 1) suggest that valence and GoF measures do differ significantly. Most relevant to the above issue is a highly significant two-way interaction between rating type and tonal stability (F=18.80, p<.001), itself modulated by a three-way interaction of rating type, tonal stability and musical experience (F=6.62, p<.002). Subsidiary ANOVAs, performed separately for musicians and non-musicians (Table 1b, c), show that while the interaction of rating type and tonal stability is significant for musicians (F=9.04, p<.001), it is not for non-musicians (F<1). A post-hoc analysis with Bonferroni corrections revealed that within musicians, GoF ratings were significantly higher than valence ratings for stable diatonic tones (corrected p<.001), but not for the other two categories (corrected p=1 and p=0.852 for unstable diatonic and chromatic tones, respectively). For musicians, then, high tonal "fit" (associated with tonal stability) does not necessarily imply highly positive valence.

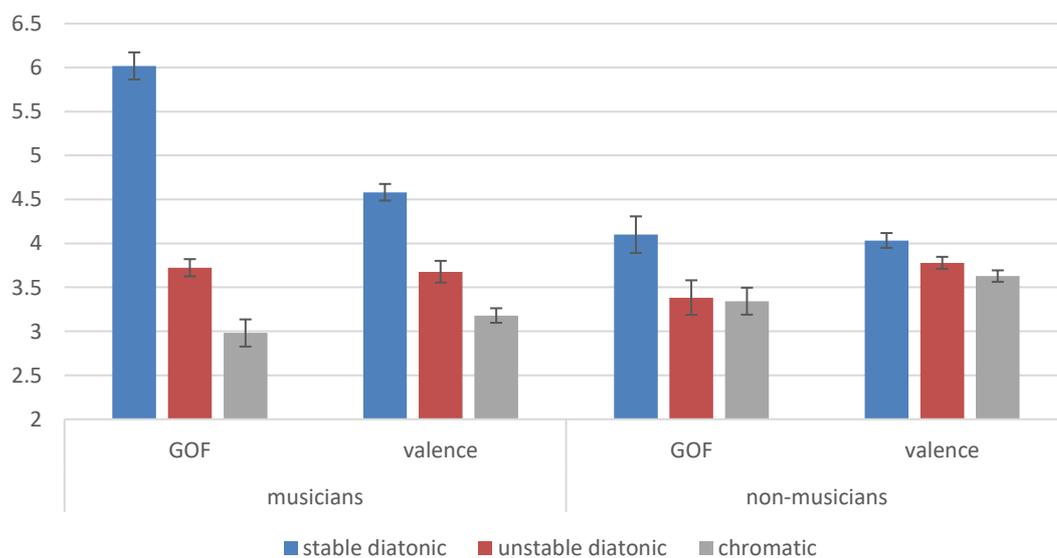

Figure 4. Mean ratings for valence (facial expressions matchings, Experiments 1a and 1b combined) and GoF (from Maimon et al., 2020) as a function of tonal stability

categories: stable diatonic (blue), unstable diatonic (red) and chromatic tones (grey), for musicians (left panel) and non-musicians (right panel). 7=highest GoF rating & happiest face, 1=lowest GoF rating & saddest face.

| | Effect | Df | Sum Squares | Mean Squares | F value | Pr(>F) |
|---|---|---|---|---|---|---|
| A. Musicians & Non-Musicians | Tonal Stability | 2 | 56.21 | 28.10 | 78.73 | <.001 |
| | Rating Type | 1 | 0.68 | 0.68 | 1.90 | 0.169 |
| | Musical Experience | 1 | 5.01 | 5.01 | 14.02 | <.001 |
| | **Tonal Stability X Rating Type** | 2 | 13.43 | 6.71 | 18.80 | **<.001** |
| | Tonal Stability X Musical Experience | 2 | 26.84 | 13.42 | 37.59 | <.001 |
| | **Rating Type X Musical Experience** | 1 | 6.60 | 6.60 | 18.50 | **<.001** |
| | **Tonal Stability X Rating Type X Musical Experience** | 2 | 4.73 | 2.36 | 6.62 | **0.002** |
| | Residuals | 282 | 100.66 | 0.36 | <1 | |
| B. Musicians | Tonal Stability | 2 | 85.74 | 42.87 | 124.03 | <.001 |
| | **Rating Type** | 1 | 2.53 | 2.53 | 7.31 | **0.008** |
| | **Tonal Stability X Rating Type** | 2 | 6.25 | 3.12 | 9.04 | **<.001** |
| | Residuals | 141 | 48.73 | 0.35 | <1 | |
| C. Non-Musicians | Tonal Stability | 2 | 5.76 | 2.88 | 7.72 | 0.001 |
| | **Rating Type** | 1 | 0.49 | 0.49 | 1.31 | **0.254** |
| | **Tonal Stability X Rating Type** | 2 | 0.47 | 0.24 | 0.63 | **0.532** |
| | Residuals | 138 | 51.48 | 0.37 | <1 | |

Table 1. Results of repeated measures ANOVAs comparing mean valence ratings (based on facial expression matching) in Experiments 1a, 1b (combined) with mean GoF ratings in Maimon et al. (2020), for both musicians and non-musicians (A), musicians (B), and non-musicians (C). Effects involving rating type (GoF vs. valence) are printed in a **bold** font.

While no significant interaction between rating type and tonal stability was observed for non-musicians, a comparison of non-musicians' ratings of tonal stability categories for valence and GoF does merit some attention. The differences between stable diatonic (scale degrees 1, 3, 5) and unstable diatonic categories (scale degrees 2, 4, 6, 7) were significant for both valence and GoF among non-musicians (Bonferroni adjusted $p=.006$ and $p=0.002$ respectively). However, while the difference between the unstable diatonic and chromatic categories was significant in non-musicians' valence matchings (adjusted $p=.037$), it did not reach significance levels in their GoF ratings (adjusted $p=1$). This may suggest a higher sensitivity among non-musicians to the emotional significance of tonality, compared to its structural significance (tonal "fit").

A second result of interest is the significant two-way interaction between musical experience and rating type ($F=18.5$, $p<.001$). Subsidiary ANOVAs revealed a significant effect of rating type for musicians ($F=7.31$, $p=0.008$), stemming from higher mean ratings for GoF compared to valence, but not for non-musicians ($F=1.31$ $p>.05$). A post-hoc analysis revealed that while musicians' GOF ratings were generally higher than non-musicians' ($p=.005$), valence ratings were the same for musicians and non-musicians ($t<1$).

These overall differences between GoF and valence ratings, particularly the higher ratings of stable diatonic tones (1, 3, 5 scale degrees) in GoF, are reflected in relationships of specific scale degrees (see Figure 5). Importantly, while GoF ratings closely match conventional tonal hierarchy for all scale degrees, valence ratings seem inconsistent with that hierarchy in several noteworthy cases. Consider, for instance, scale degrees 3 (mediant) and 4 (subdominant) – the former considered a stable degree, the latter an instable, tensive one, typically resolved toward it. While GoF ratings clearly reflect this hierarchy, valence ratings reverse it (Figures 5 A-D). A comparable situation emerges between the stable 5[th] (dominant) and instable 6[th] (submediant) degrees in minor: for GoF, the former is rated vastly higher, while for valence, mean ratings are almost identical (Figures 5B, 5D).

A divergence of particular interest between GOF and valence ratings involves the diatonic and the raised (3#, so-called the "Picardy third") 3$^{rd}$ degrees in the minor mode, particularly for musicians (see Figure 5B). While diatonic scale-degree 3 is a stable tone, the Picardy third is a chromatic tone. However, the Picardy third degree has traditionally been used as a part of a cadential major tonic chord in minor mode pieces (particularly in sacred music), often associated with emotionally positive ending of a somber piece (Kivy, 1999, p. 289; Rushton, 2001). This twofold relationship is conveyed by the different GoF and valence ratings of the Picardy 3$^{rd}$, compared to the diatonic 3$^{rd}$. For GoF, ratings of the diatonic 3$^{rd}$ are vastly higher, reflecting theoretical tonal hierarchy (M$\Delta$=1.81, SD=1.68). In contrast, mean valence ratings of the two degrees are considerably closer, as rating for the diatonic 3$^{rd}$ is relatively lower compared to GoF, while that of the Picardy 3$^{rd}$ is relatively higher (M$\Delta$=0.46 SD=1.43). Note also that the Picardy 3rd, while rated as low as other chromatic tones for GoF, is rated as high as diatonic tones for valence (Figure 5B). Between-participants t-tests indicate that the difference between the diatonic and the Picardy 3$^{rd}$ was indeed significantly larger in GoF ratings compared to valence matchings within musicians ($p$<.001), though not within non-musicians ($p$=.336). Notably, while 95% of the musicians in Maimon et al. (2020) -- 19 of 20 -- rated the diatonic 3rd as more fitting than the Picardy 3rd in the GOF experiment, only 30% of musicians in Experiment 1a (6 of 20) chose happier faces for the diatonic compared to the Picardy 3rd. Thus, though the Picardy 3$^{rd}$ degree seems to be perceived as unfitting its tonal context, it is certainly not perceived as particularly gloomy: tonal valence seems to be divorced, in this case, from tonal fit.

To sum up: comparison of valence and GoF ratings suggests that perceived "tonal valence" – the emotional valence associated with tonal scale-degrees, as investigated in the present study – is distinct of perceived tonal fit, as conveyed by GoF ratings. We revealed general distinctions between these two dimensions, particularly the higher ratings of stable diatonic degrees for GoF, and highlighted their significance for specific scale degrees such as the Picardy third. Notably, GoF-valence distinctions were revealed mainly for musicians, suggesting that musical training enhances the sensitivity to both similarities and disparities between emotional and structural features of music.

**A**

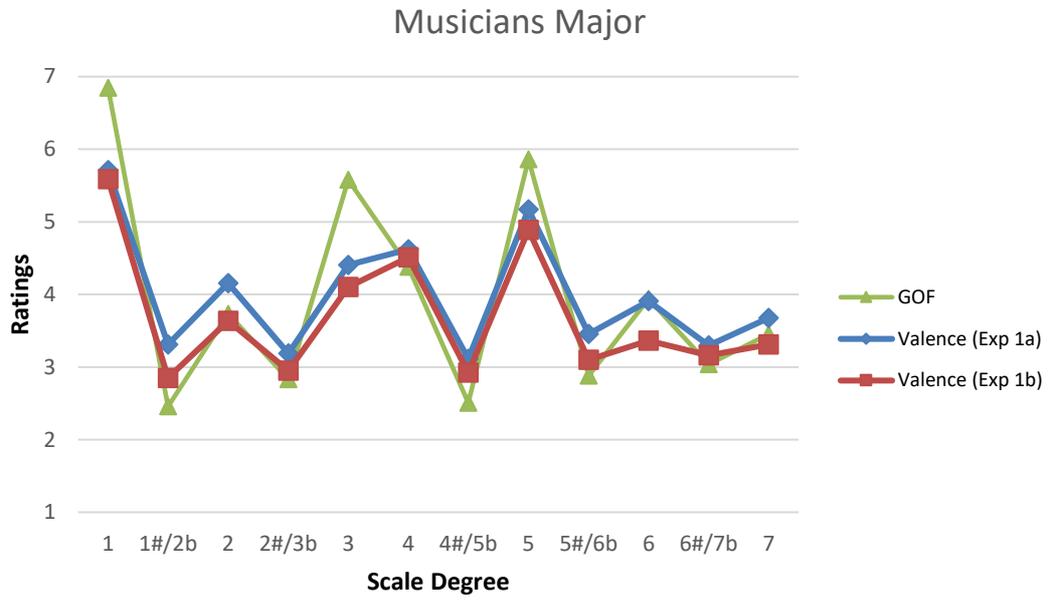

**B**

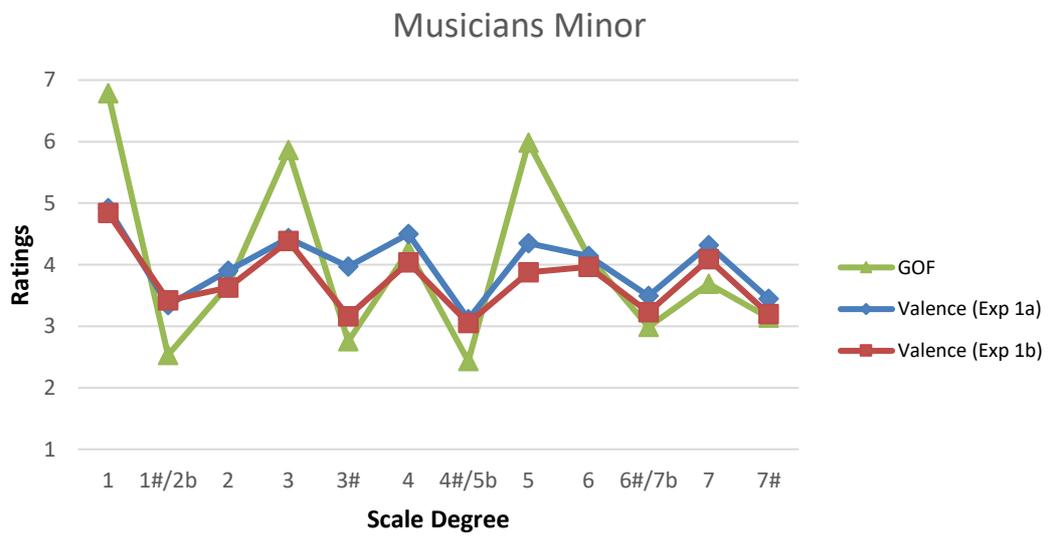

**C**

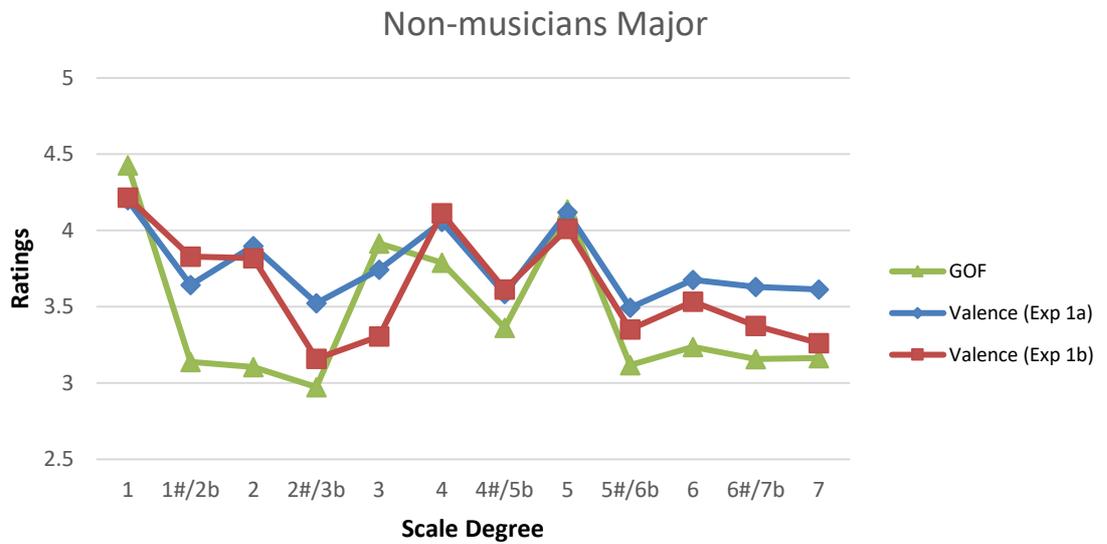

**D**

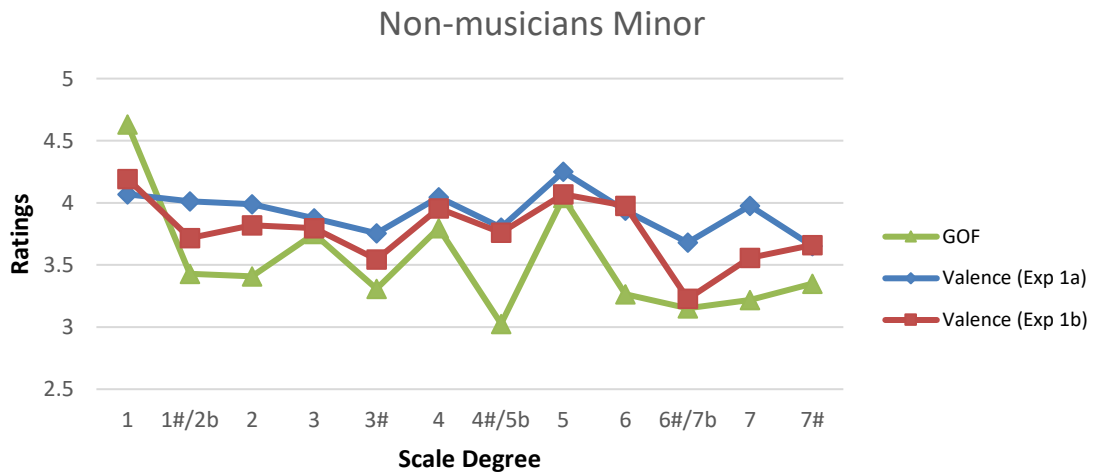

Figures 5. Mean valence (facial expressions) matchings in experiment 1a (blue), experiment 1b (red) and GoF ratings in Maimon et al 2020 (green), for musicians in the major mode (A), musicians in the minor mode (B), non-musicians in the major mode (C), and non-musicians in the minor mode (D). Y-axis values for GOF ratings range from 1(least fitting) to 7 (most fitting); for valence (facial expression matchings): 1- saddest expression, 7- happiest expression.

# Experiment 2

The modified probe-tone paradigm used in Experiment 1 relies on *explicit* associations between tonal stability and emotional valence, as the assignments of emotional valence ratings are based on participants' conscious introspection. To examine whether similar correspondence may also be evoked implicitly and involuntary, affecting faster, task-irrelevant responses , a different experimental paradigm is needed.

In Experiment 2, we examined associations between tonal stability and emotional valence by using a variant of the Implicit Association Test (IAT; Greenwald, McGhee, & Schwartz, 1998). The IAT was originally designed to assess implicit attitudes – evaluations that occur unintentionally and unconsciously – towards certain population groups. It was later adapted by Parise and Spence (2012) to gauge associations between perceptual dimensions in different modalities indirectly. Experiment 2 is based on that cross-modal version of the paradigm.

On each trial of our IAT experiment, participants were presented with one of four kinds of stimuli, two auditory and two visual. The auditory stimuli were a tonally stable progression, [A]stable, and a tonally unstable progression [A]unstable (i.e., a key-establishing chord sequence followed by a stable or an unstable tone in that key, respectively). The visual stimuli were a happy and a sad face, [V]Happy, [V]Sad, presented on a computer screen. On each trial, participants were asked to categorize a single stimulus (auditory or visual) as quickly and accurately as possible. In each block of trials, two stimuli – one auditory and one visual – were assigned the same response key. Thus, throughout the block, participants used only two keys, each key paired with one value of the auditory stimuli and one value of the visual stimuli. There were two types of blocks. In congruent blocks, the hypothetically compatible auditory and visual stimuli — [A]stable and [V]Happy, [A]instable and [V] Sad — were assigned the same response keys, while in incongruent blocks, incompatible stimuli — [A]stable and [V]Sad, [A]instable and [V]Happy— were assigned to the same response keys. We expected performance in congruent blocks to be better (faster and more accurate) than in incongruent blocks. Specifically, we reasoned that if tonality and emotional valence are associated involuntarily and implicitly in the same way as they are associated voluntarily and explicitly (Experiments 1A and B), then using the same key for [A]stable and [V]Happy stimuli and for [A]unstable and [V]Sad stimuli should be easier than using different keys.

As noted above, the IAT paradigm does not require any explicit judgement of the association of the auditory and visual stimuli, that association being orthogonal to the task. Thus, effects of congruence between tonal stability and emotional faces on the dependent variables (RT and accuracy) may reflect an automatic, involuntary association between the two dimensions.

### Materials and Methods

#### Sample size selection

Based on the main effect of Parise and Spence (2012) (Exp. 3, pitch height and size), we calculated the sample size required to observe a significant compatibility effect between tonal stability and visual brightness. We conducted this analysis with G*Power (Erdfelder *et al.*, 2013) using an alpha of 0.05, and a power of 0.80. We found the minimum required sample size to be four participants. As we examined possible interactions of this effect with musical training (musicians vs non-musicians) and modality (auditory vs visual), the minimum sample size was 16.

#### Participants

Participants (n=40, 20 musicians) were the same participants as in experiment 1a.

The apparatus was the same as in Experiment 1a.

*Stimuli*

*Auditory stimuli*. Audio examples and graphical representations of the auditory stimuli are available in supplementary material A. The auditory stimuli consisted of a half-cadence (IV-V$^4_6$-$^3_5$) followed by either the tonic note (a tonally stable stimulus) or the raised subdominant (a tonally unstable chromatic note), as illustrated in Figure 6. Each stimulus was presented in two keys (C major, D-flat major), in a piano timbre (generated by Notion 6 music notation software).[3] To minimize possible STM effects, the cadence (played in a three-voice texture) was designed such that neither of the final tones (tonic, raised 4th) was included in the chords preceding it. The final tone was octave-doubled, and the pitch direction (up/down) between the last chord in the cadence and the following tone was controlled.

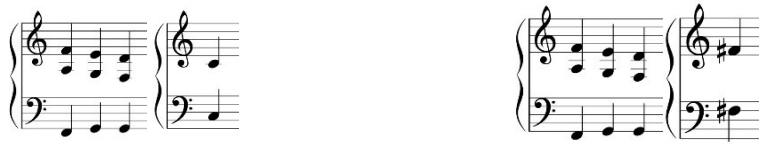

Figure 6. Examples of the stable (left) and unstable (right) auditory stimuli used in Experiment 2.

*Visual stimuli*. The visual stimuli consisted of eight faces taken from stimuli in experiment 1. We used photographs of veridical faces taken from the Data Organization of BU-3DFE Database. In the present experiment, we selected four Caucasian individuals (two women), the third level of sadness and the third level of happiness for each of the selected individuals.

---

3. A comparable experiment using minor mode stimuli was planned but not performed due to the COVID-19 pandemic.

*Procedure*

On each trial, participants were presented with a unimodal stimulus (either auditory or visual). They were asked to classify it using one of two keys (K or D). For the visual stimuli, one key was assigned to the happy face, and another to the sad face. For the auditory stimuli, one key was assigned to a stable progression, and the other to an unstable progression. Note that the experimenter did not mention any of these adjectives (happy, sad, stable or unstable) but instead referred to the visual and auditory stimuli as type K and type D according to the participant's key assignment in the practice blocks. The experiment consisted of 24 blocks. Each block consisted of 28 consecutive trials. For each block, only one pair of faces (the happy and sad version of the same individual's face) was selected and used throughout the block. The first four trials served as practice trials and included one stimulus of each type (auditory stable, auditory unstable, happy face and sad face). In the following 24 experimental trials, the two faces were presented 6 times each, and the four auditory stimuli (stable and unstable in C and C# keys) were presented 3 times each, all randomly mixed. Each of the four possible stimulus-key pairings (2 congruent and 2 incongruent pairings) was presented in different blocks. There were 6 blocks for each pairing, resulting in 24 experimental blocks presented in randomly mixed order. Thus, in half of the blocks, the response pairing was congruent (i.e., the same response was associated with the stable auditory stimuli and happy face and with the unstable auditory stimuli and the sad face), whereas in the remaining half of the blocks, the response pairing was incongruent (i.e., the same response was associated with the unstable auditory stimuli and the happy face and with the stable auditory stimuli and the sad face).

*Results*

The data from one non-musician participant was excluded because his accuracy rate was lower than the group's mean by more than 3 standard deviations (60% vs. M=85.33%, SD =8.54% for non-musicians).

A repeated-measures ANOVA was conducted, with musical training (musician vs. non-musician) as a between-participants variable, compatibility (congruent/incongruent) and modality (auditory/visual) as within-participant variables, and mean accuracy rates and reaction times for correct responses as the dependent variables. The mean reaction times and accuracy rates in Experiment 2 are presented in Figure 7.

*Reaction times*. The main effect of compatibility was significant, suggesting that participants were faster on congruent than on incongruent blocks, $F(1,37)=17.28$, $p<0.001$, $\eta_p^2=.32$. This effect interacted with modality, $F(1,37)=10$, $p=0.003$, $\eta_p^2=.21$, indicating that the compatibility effect was larger for the visual than for the auditory modality, but was significant for both, $F(1,37)=23.73$, $p<.001$ $\eta_p^2=.39$ and $F(1,37)=6.93$, $p=.01$ $\eta_p^2=.16$ for the visual and auditory modality respectively. There were no other significant effects. In particular, the compatibility effect was not involved in any interaction with either musical experience or modality, all Fs<1.

*Accuracy rates*. The main effect of compatibility was significant, indicating that participants were faster on congruent than on incongruent blocks, $F(1,37)=43.12$, $p<0.001$, $\eta_p^2=.54$. The main effect of musical experience was also significant, $F(1,37)=11.75$, $p=0.002$, $\eta_p^2=.24$, but this effect was modulated by a significant interaction with modality , F(1,37)=8.64, p=0.006, $\eta_p^2$=.19: while musicians were more accurate in the auditory modality, non-musicians were more accurate in the visual modality. There were no other significant effects. Again, the compatibility effect was not involved in any interaction with either musical experience or modality, all Fs<1.

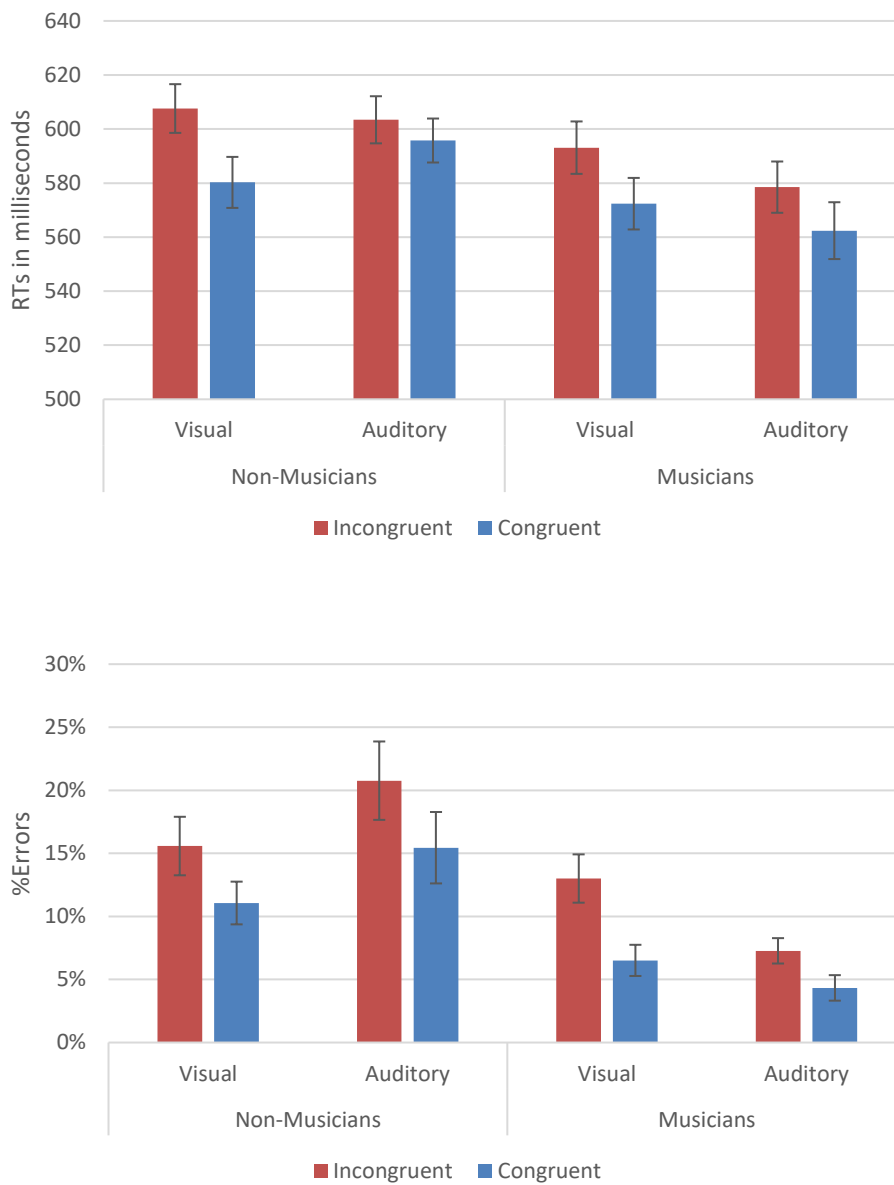

Figure 7. Mean reaction times (upper graph) and error rates (lower graph) in Experiment 2, for congruent (blue) and incongruent (red) blocks, by modality (visual and auditory) for non-musicians (left) and for musicians (right).

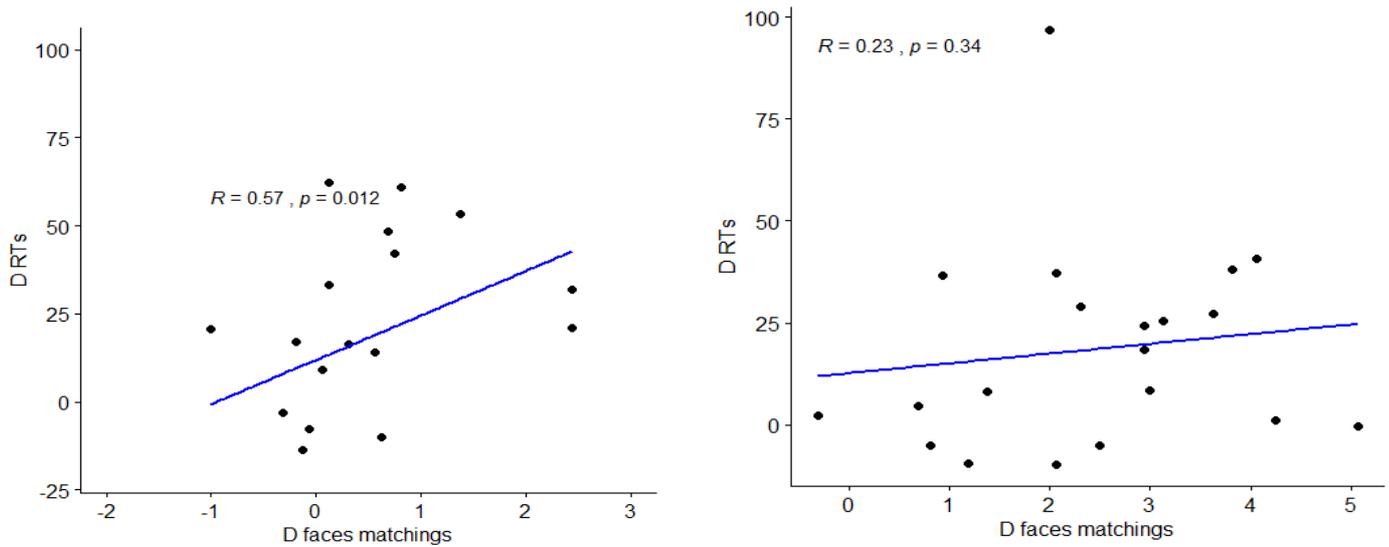

Figure 8. non-musicians' (left) and musicians' (right) within-participants correlations between the delta between $1^{st}$ vs. $4\#^{th}$ faces matchings in Exp 1A (x axis) and delta between RTs in compatible vs. incompatible blocks in Experiment 2 (y axis).

*Within participants' correlations (Experiments 1A, 2): an exploratory analysis*

Since the same participants took part in Experiments 1A and 2, we examined the within-participant rank correlations between comparable data in the two experiments in an exploratory investigation. For Experiment 2, we considered the mean difference in reaction times between compatible and incompatible blocks; for Experiment 1A – the mean difference between valence ratings of the same degrees used as the stable and unstable stimuli in Exp 2 (e.g. the tonic note, and the raised $4^{th}$). For both groups together (N=39), the mean within-participant rank correlation between these values was not significant, Spearman rho=.272, p=.094 (see figure 8). Separate examination of musicians' (N=20) and non-musicians' (N=19) data revealed that while musicians' rank correlation did not reach significance, Spearman rho=.226, p=.337, for non-musicians it was significant, Spearman rho=.566, p=.012.

*Discussion*

Experiment 2 complements Experiment 1 by examining implicit, involuntary associations, rather than explicit associations, of tonal stability with emotional valence. Results suggest that tonal stability and emotional valence are associated involuntarily, as stimuli in compatible blocks (where stable tonal progressions were matched with happy facial expressions, and instable progression – with sad faces) were processed faster and more accurately than stimuli in incompatible blocks. Notably, musical experience did not interact with these compatibility effects, suggesting that unlike the case with explicit associations, (Experiment 1), musical expertise did not affect involuntary, speeded associations of tonal stability and valence. Complementarily, non-musicians' results significantly correlated with their results in Experiment 1a, while musicians' results did not. Together, these results suggest that results of Experiments 1 and 2 present different processing mechanisms: a largely non-conceptual mechanism, possibly used by non-musicians in both experiments and by musicians in Experiment 2, and a reflective, conceptually based mechanism, used by musicians in Experiment 1.

## General discussion

This study investigated how listeners, musicians and non-musicians, associate tonal hierarchy and emotional valence. In an explicit task (Experiments 1a, 1b), participants matched each of the 12 chromatic notes, following a major or minor tonal context, with facial expressions ranging from negative to positive in emotional valence. In a speeded task (Experiment 2), applying a variant of the Implicit Association Test (IAT), they rapidly responded, following a priming tonal context, to hypothetically congruent or incongruent combinations of tones and facial expressions.

Three main findings of the study may be underscored. First, we established that tonal stability is associated with perceived emotional valence, such that more stable tones are associated with more positive valence. Importantly, while associations of tonality and emotion have been previously studied empirically, this is the first study that maps the entire gamut of diatonic and chromatic melodic scale degrees, in both major and minor contexts, to a perceived emotional dimension (valence).

Second, the results suggest in several ways that the association of tonal stability and valence relies on two distinct mechanisms: one mediated by conceptual musical knowledge and conscious decisional processes, the other – by non-conceptual and more automatic processes. (1) The association between tonality and perceived emotional valence was observed both in musicians, who possess relevant conceptual knowledge (e.g., the ability to identify particular scale-degrees), and in non-musicians, who do not possess such knowledge. (2) The association emerged both in an explicit matching task, in which consciously mediated decisional processes are allowed to play a role (Experiments 1a, 1b), and in an implicit speeded task (IAT, Experiment 2), thought to rely much less on such processes. (3) We found musical training to modulate the effect of tonal stability in the explicit task, with stronger effects in musicians than in non-musicians, but not in the speeded task, where both groups showed similar effects. Thus, at the lower level of processing, which is presumably less influenced by top-down cognitive control, the emotional associations of tonal stability appeared to be similar in musicians and non-musicians. Consistent with this finding, an initial exploratory correlation between the effects was significant within the musically untrained participants (who presumably relied only on non-conceptual knowledge in both tasks) but not in musicians (who could rely on conceptual knowledge of music theory in the explicit matching task but not in the speeded IAT task).

Third, given the robust association of tonal stability and perceived emotional valence reported above, one may ask whether investigating the latter might in effect be identical to investigating the former: whether "tonal valence" – the perceived valence of a tone within a tonal framework, examined here – can be merely reduced to a sense of how well that tone fits that framework, that is, to the well-established measure of "Goodness of Fit" (Krumhansl, 1990; Krumhansl & Kessler, 1982). The present study suggests that this is not the case – that responding to the way a tone *feels* in a tonal context differs from responding to how it *fits* that context. A comparison of the valence ratings in the present experiments (1a, 1b) with goodness-of-fit ratings of the same stimuli (Maimon et al., 2020) reveals significant differences between these two measures. In particular, GoF ratings of stable scale degrees are considerably higher than their valence ratings, and specific scale degrees (such as the raised, "Picardy" third degree in minor) are rated in markedly different ways for GoF and valence. These differences suggest that perceived tonal valence and perceived tonal stability are distinct, though related dimensions. Notably, however, these differences between structural and emotional dimensions may apply mostly to the musically-trained, where distinctions between GoF and valence measures proved to be highly significant.

While tonal hierarchy associated with valence, mode (major or minor), often referred to as a principal determinant of emotional valence in music (e.g., Parncutt, 2014) did not: effects of mode on facial expression matchings were not found in either Experiment 1a or 1b. This null effect could stem from the fact that major and minor stimuli were presented in separate blocks. As a result, participants may have matched stimuli mainly relative to other stimuli in the same block, which would wipe out the expected differences between the major and minor modes. Note, however, that the association of tonal stability and emotional valence was stronger in the major than in the minor mode, a difference likely to reflect the weaker distinction between diatonic and chromatic degrees in minor (Meyer, 1956; Parncutt, 2014). An effect of mode was also revealed indirectly in the relatively high ratings of the Picardy 3rd (see above).

Beyond its findings and their implications, this study proposes a new experimental design, adapting two well-established behavioral paradigms, Krumhansl's probe-tone method (e.g., Krumhansl & Kessler, 1982), and the Implicit Association Test (IAT), to cross-modal stimuli (Parise & Spence, 2012). Both methods could be applied to investigating diverse connotative meanings of tonal structure. In our modified probe-tone test, rather than rating probe-tones for goodness-of-fit, participants matched each probe with one of a graded series of visual stimuli (here, facial expressions gradually ranging from "happy" to "sad"). Such matchings, converted into a numerical scale, were examined vis-a-vis other scalable tonal features, particularly GOF ratings and scale-degrees' stability ranking based on tonal theory. Importantly, this novel paradigm is not limited to examining emotional valence but may be applied to investigate diverse cross-domain connotations of tonal structure (see Eitan, 2017, for a related discussion). In other studies using the two paradigms applied here, our group established a range of cross-modal connotations of the tonal scheme, associating tonal stability and closure with visual brightness (Maimon, et al., 2020), physical size, or spatial location (Maimon, Lamy, & Eitan, 2019).

Our adaptation of the Implicit Association Test complements the use of the explicit probe-tone method and could likewise be extended to examine other connotations of tonal hierarchy. The task in the IAT paradigm, unlike that of the probe-tone method, does not require reflective judgement of the association between dimensions of interest (here, tonal stability and facial expression), and such association – conscious or unconscious – is orthogonal to the task. Hence, the correspondences between tonal stability and facial expression implied by results of the IAT tests need not be based on a voluntary or conscious decisional process alone, but might be encapsulated in involuntary and more automatic processes, associated with faster responses to task irrelevant dimensions. It should be noted that the implicit and automatic nature of responses in the IAT paradigm has been called into question, based on claims that in most IATs the purpose of testing is not highly concealed (e.g., Fiedler et al., 2006), that it is possible to intentionally influence IAT's scores with deliberate thinking (e.g., Dasgupta & Greenwald, 2001) and even fake results upon request (Fiedler & Bluemke, 2005). However, these criticisms apply mainly to IAT testing easily-recognized attributes, laden with social significance. They are less likely to apply to the perceptual associations examined here, especially considering that 50% of our participants (non-musicians) had no conceptual knowledge about the critical auditory feature (tonal stability).

Applying both explicit (matching) and implicit (IAT) paradigms together, while comparing participants with different levels of expertise (musicians vs. non-musicians) may allow a full orthogonal approach, dissociating explicit strategies, based on conscious reflection, faster involuntary processes, and conceptual expert knowledge. This approach may be applied to diverse connotations, as already done in Maimon et al. (2019, 2020), thus contributing to a comprehensive empirical investigation of expressive and cross-modal associations of musical tonality, covering diverse connotative domains and different levels of processing. Such empirical examination of the connotative meanings of tonality may help elucidate a central issue in musical discourse: how an abstract sound structure, a musical syntax, associates in our minds with the "real" non-auditory world of perceptions, emotions, and actions.

Acknowledgements: We thank Roni Granot, Liad Mudrik and Yeshayahu Shen for very useful discussions, and Inbal Gur-Arie for her help in performing the experiments. Special thanks to Miriam Furst-Yust of TAU's School of Electrical Engineering for valuable technical assistance.

Funding: This research is supported by an Israel Science Foundation grant (ISF 1920/16) to Zohar Eitan

Declaration of interest statement: None

**References**

Aarden, B. J. (2003). *Dynamic melodic expectancy* (Doctoral dissertation, The Ohio State University).

Albrecht, J., & Shanahan, D. (2013). The use of large corpora to train a new type of key-finding algorithm: An improved treatment of the minor mode. *Music Perception: An Interdisciplinary Journal*, *31*(1), 59-67.

Arthur, C. (2018). A perceptual study of scale-degree qualia in context. *Music Perception: An Interdisciplinary Journal*, *35*(3), 295-314/

Bigand, E., Tillmann, B., & Poulin-Charronnat, B. N. D. (2006). A module for syntactic processing in music?. *Trends in cognitive sciences*, *10*(5), 195-196.

Blair, I. V., Ma, J. E., & Lenton, A. P. (2001). Imagining stereotypes away: the moderation of implicit stereotypes through mental imagery. *Journal of personality and social psychology*, *81*(5), 828.

Carlsen, J. C. (1981). Some factors which influence melodic expectancy. *Psychomusicology: A Journal of Research in Music Cognition*, *1*(1), 12.

Cheung, V. K., Harrison, P. M., Meyer, L., Pearce, M. T., Haynes, J. D., & Koelsch, S. (2019). Uncertainty and surprise jointly predict musical pleasure and amygdala, hippocampus, and auditory cortex activity. *Current Biology*, *29*(23), 4084-4092.

Cooke, D. (1959). *The Language of Music*. London and New York: OUP.

Cuddy, L. L., & Lunney, C. A. (1995). Expectancies generated by melodic intervals: Perceptual judgments of melodic continuity. *Perception & Psychophysics*, *57*(4), 451-462.

Deyoe, N., & Bellman, J. (2005). Joseph Haydn: First Movements of Six Unnamed Minor Symphonies.

Erdfelder, E., Faul, F., & Buchner, A. (1996). GPOWER: A general power analysis program. *Behavior research methods, instruments, & computers*, *28*(1), 1-11.

Probabilistic models of expectation violation predict psychophysiological emotional responses to live concert music.

Eitan, Z. (2017). Cross-modal correspondences, in: *The Routledge Companion to Music Cognition*, R. Ashley and R. Timmers (Eds), pp. 213–224, Routledge, New York, NY, USA.

Eitan, Z., Ben-Haim, M. S., & Margulis, E. H. (2017). Implicit absolute pitch representation affects basic tonal perception. *Music Perception: An Interdisciplinary Journal*, *34*(5), 569-584.

Fétis, F. J. (1849). *Traité complet de la théorie et de la pratique de l'harmonie: contenant la doctrine de la science et de l'art* (Vol. 4). Brandus.

Fiedler, K., Messner, C., & Bluemke, M. (2006). Unresolved problems with the "I", the "A", and the "T": A logical and psychometric critique of the Implicit Association Test (IAT). *European Review of Social Psychology*, *17*(1), 74-147.

Fiedler, K., & Bluemke, M. (2005). Faking the IAT: Aided and unaided response control on the Implicit Association Tests. *Basic and Applied Social Psychology*, *27*(4), 307-316.

Granot, R., & Donchin, E. (2002). Do Re Mi Fa Sol La Ti——Constraints, Congruity, and Musical Training: An Event-Related Brain Potentials Study of Musical Expectancies. *Music Perception*, *19*(4), 487-528.

Greenwald, A. G., & Nosek, B. A. (2001). Health of the Implicit Association Test at age 3.

Greenwald, A. G., McGhee, D. E., & Schwartz, J. L. (1998). Measuring individual differences in implicit cognition: the implicit association test. *Journal of personality and social psychology*, *74*(6), 1464.

Huron, D. (2006). *Sweet Anticipation: Music and the Psychology of Expectation*. MIT Press, Cambridge, MA, USA.

Juslin, P. N., Harmat, L., & Eerola, T. (2014). What makes music emotionally significant? Exploring the underlying mechanisms. *Psychology of Music*, *42*(4), 599-623.

Juslin, P. N., & Vastfjall, D. (2008). Emotional responses to music: The need to consider underlying mechanisms. *Behavioral and brain sciences*, *31*(5), 559.

Kallinen, K., & Ravaja, N. (2006). Emotion perceived and emotion felt: Same and different. *Musicae Scientiae*, *10*(2), 191-213.

Kawakami, A., Furukawa, K., Katahira, K., Kamiyama, K., & Okanoya, K. (2012). Relations between musical structures and perceived and felt emotions. *Music Perception: An Interdisciplinary Journal*, *30*(4), 407-417.

Kivy, P. (1999). *Osmin's Rage: Philosophical Reflections on Opera, Drama, and Text, with a New Final Chapter*. Ithaca and London: Cornell University Press.

Koelsch, S., Fritz, T., & Schlaug, G. (2008). Amygdala activity can be modulated by unexpected chord functions during music listening. *Neuroreport*, *19*(18), 1815-1819.

Krumhansl, C. L. (2004). The cognition of tonality–as we know it today. *Journal of New Music Research*, *33*(3), 253-268.

Krumhansl, C. L., & Kessler, E. J. (1982). Tracing the dynamic changes in perceived tonal organization in a spatial representation of musical keys. *Psychological review*, *89*(4), 334.

Krumhansl, C. L., & Schmuckler, M. (1990). A key-finding algorithm based on tonal hierarchies. *Cognitive Foundations of Musical Pitch*, 77-110.

Maimon, N., Lamy, D. and Eitan, Z. (2019). Associations of tonal stability and visual space: explicit and implicit measures. Talk presented at the 21th conference of the European Society for Cognitive Psychology, Tenerife, SP.

Maimon, N. B., Lamy, D., & Eitan, Z. (2020). Crossmodal Correspondence Between Tonal Hierarchy and Visual Brightness: Associating Syntactic Structure and Perceptual Dimensions Across Modalities. *Multisensory Research*, *1*(aop), 1-32.

Meyer, L. B. (1956). *Emotion and Meaning in Music*. University of Chicago Press, Chicago, IL, USA.Miceli, M., & Castelfranchi, C. (2014). *Expectancy and emotion*. OUP Oxford.

Parise, C. V., & Spence, C. (2012). Audiovisual crossmodal correspondences and sound symbolism: a study using the implicit association test. *Experimental Brain Research*, *220*(3), 319-333.

Parncutt, R. (2014). The emotional connotations of major versus minor tonality: One or more origins?. *Musicae Scientiae*, *18*(3), 324-353.

Proulx, T., Inzlicht, M., & Harmon-Jones, E. (2012). Understanding all inconsistency compensation as a palliative response to violated expectations. *Trends in cognitive sciences*, *16*(5), 285-291.

Rameau, J. P. (1722). *Traité de l'Harmonie Reduite à ses Principes Naturels: Divisé en Quatre Livres.* Jean-Baptiste-Christophe Ballard, Paris, France.

Rusthon, J. (2001). Tierce de Picardie [Picardy 3rd]. *Groves Music Online*, https://doi.org/10.1093/gmo/9781561592630.article.27946

Schoenberg, A. (1978). *Theory of harmony*. University of California Press, Berkeley, Los Angeles, USA.

Schmuckler, M. A. (1989). Expectation in music: Investigation of melodic and harmonic processes. *Music Perception*, *7*(2), 109-149.

Schmuckler, M. A. (2016). Tonality and contour in melodic processing, in: S. Hallam, I. Cross, & M. Thaut (Eds.), Oxford library of psychology. *The Oxford handbook of music psychology* (pp. 143–165). Oxford University Press. Oxford, UK.

Shanahan, D. (2017). Musical Structure: Tonality, Melody, Harmonicity, and Counterpoint, in: *The Routledge Companion to Music Cognition* (pp. 141-151). Routledge, NY, NY, USA.

Shepard, R. N. (1964). Circularity in judgments of relative pitch. *The journal of the acoustical society of America*, *36*(12), 2346-2353.

Sloboda, J. A. (1991). Music structure and emotional response: Some empirical findings. *Psychology of music*, *19*(2), 110-120.

Smit, E. A., Dobrowohl, F. A., Schaal, N. K., Milne, A. J., & Herff, S. A. (2020). Perceived Emotions of Harmonic Cadences. *Music & Science*, *3*, 2059204320938635.

Steinbeis, N., Koelsch, S., & Sloboda, J. A. (2006). The role of harmonic expectancy violations in musical emotions: Evidence from subjective, physiological, and neural responses. *Journal of cognitive neuroscience*, *18*(8), 1380-1393.

Zatorre, R. J., & Salimpoor, V. N. (2013). From perception to pleasure: music and its neural substrates. *Proceedings of the National Academy of Sciences*, *110*(Supplement 2), 10430-10437.